\documentclass[a4paper]{article}
\usepackage{RR}
\usepackage{hyperref}
\usepackage{amsmath,amssymb,graphicx,subfigure}
\usepackage{pslatex,xspace}
\usepackage{rotating}
\usepackage{amstext}
\usepackage{amsopn}

\RRdate{Juillet 2007}
\RRauthor{
Fr\'ed\'eric Havet\thanks[sfn]{MASCOTTE, I3S (CNRS--UNSA) -- INRIA, 2004
Route des Lucioles, BP 93, 06902 Sophia Antipolis Cedex, France. E-mails: {\tt
\{fhavet,sereni\}@sophia.inria.fr}}%
  \and
Jean-S\'ebastien Sereni\thanksref{sfn}%
  \and
Riste \v{S}krekovski\thanks{Department of Mathematics, University of Ljubljana,
Jedranska 19, 1111 Ljubljana, Slovenia. Supported in part by
Ministry of Science and Technology of Slovenia, Research Program
P1-0297. E-mail:
\texttt{bluesky2high@yahoo.com}}
}
\authorhead{Havet, Sereni \& \v{S}krekovski}
\RRtitle{Coloration $3$-faciale des graphes planaires}
\RRetitle{$3$-facial colouring of plane graphs}
\titlehead{$3$-facial colouring of plane graphs}

\renewcommand\l{\ensuremath{\ell}\xspace}

\RRresume{Un graphe planaire est \l-facialement $k$-colorable
s'il existe une coloration de ses sommets avec $k$ couleurs
telle que les sommets reli\'es par un chemin facial
de longueur au plus \l soient color\'es diff\'eremment.
Nous d\'emontrons que tout graphe planaire est $3$-facialement
$11$-colorable. Par cons\'equent, tout graphe planaire $2$-connexe
dont toutes les faces ont taille au plus $7$ est cycliquement $11$-colorable.
Ces deux bornes sont \`a une couleur pr\`es celles propos\'ees par la
conjecture $3\l+1$ et la conjecture cyclique.
}
\RRabstract{%
A plane graph is \l-facially $k$-colourable
if its vertices can be coloured with $k$ colours such that any two
distinct vertices on a facial segment of length at most \l are coloured
differently.
We prove that every plane graph is $3$-facially $11$-colourable.
As a consequence, we derive that every $2$-connected plane graph
with maximum face-size at most $7$ is cyclically $11$-colourable.
These two bounds are for one off from those that are proposed by
the $(3\l+1)$-Conjecture and the Cyclic Conjecture.
}
\RRmotcle{coloration faciale, coloration cyclique,  
graphes planaires} 
\RRkeyword{facial colouring, cyclic colouring, planar graphs}
\RRprojet{Mascotte}  
\RRtheme{\THCom} 
\URSophia 
\begin{document}
\makeRR   

\newtheorem{problem}{Problem}
\newtheorem{proposition}{Proposition}
\newtheorem{theorem}{Theorem}
\newtheorem{conjecture}{Conjecture}
\newtheorem{corollary}{Corollary}
\newtheorem{lemma}{Lemma}
\newtheorem{claim}{Claim}

\newenvironment{proof}[1][]{\par \noindent {\bf Proof#1}.\ }{\hfill$\Box$
\par \vspace{11pt}}

\newcommand{\dang}{\ensuremath{{\tt dgs}}\xspace}
\newcommand{\x}{\ensuremath{{\tt fce}}\xspace}
\newcommand{\y}{\ensuremath{{\tt sfe}}\xspace}
\newcommand{\z}{\ensuremath{{\tt bad}}\xspace}
\newcommand{\zz}{\ensuremath{{\tt vbd}}\xspace}
\newcommand\ch{{\rm ch}}

\renewcommand{\theenumi}{(\roman{enumi})}
\renewcommand{\labelenumi}{\theenumi}

\renewcommand{\thesubfigure}{}

\section{Introduction}\label{sec:int}
 The concept of
facial colourings, introduced in~\cite{KMS05}, extends the well-known
concept of cyclic colourings.
A {\it facial segment} of a plane graph $G$ is a sequence of vertices in the order obtained
when traversing a part of the boundary of a face. The {\it length} of a
facial segment is its number of edges.
Two vertices $u$ and $v$ of $G$ are \l-{\it facially adjacent},
if there exists a facial segment of length at most \l between them.
An {\it \l-facial colouring} of $G$ is a function which assigns a colour
to each vertex of $G$ such that any two distinct \l-facially adjacent vertices are
assigned distinct colours. A graph admitting an \l-facial colouring
with $k$ colours is called {\it \l-facially $k$-colourable}.

\begin{figure}[htbp]
\centering
\input{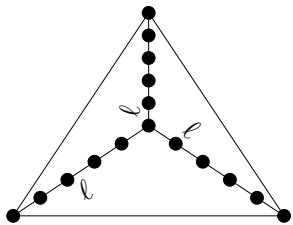}
\caption{The plane graph $G_\l=(V,E)$: each thread represents a
path of length \l. The graph $G_\l$ is not \l-facially $3\l$-colourable: every
two vertices are \l-facially adjacent, therefore any \l-facial colouring
must use $|V|=3\l+1$ colours.}
\label{fig:tight}
\end{figure}

The following conjecture, called $(3\l+1)$-Conjecture, is proposed
in~\cite{KMS05}:
\begin{conjecture}[Kr\'al', Madaras and
\v{S}krekovski]\label{conj:kms}
Every plane graph is \l-facially colourable with $3\l+1$ colours.
\end{conjecture}
Observe that the bound offered by Conjecture~\ref{conj:kms} is tight: as
shown by Figure~\ref{fig:tight}, for every $\l\ge1$, there exists a
plane graph which is not $\l$-facially $3\l$-colourable.

Conjecture~\ref{conj:kms} can be considered
 as a counterpart for
\l-facial colouring of the following famous conjecture by Ore and
Plummer~\cite{OrPl69} concerning the cyclic colouring. A plane graph
$G$ is said to be {\it cyclically $k$-colourable}, if it admits a
vertex colouring with $k$ colours such that any pair of vertices
incident to a same face are assigned distinct colours.
\begin{conjecture}[Ore and Plummer]\label{conj:op}
Every plane graph is cyclically
$\left\lfloor\frac{3\Delta^*}{2}\right\rfloor$-colourable, where
$\Delta^*$ denotes the size of a biggest face of $G$.
\end{conjecture}
Note that Conjecture~\ref{conj:kms} implies
Conjecture~\ref{conj:op} for odd values of $\Delta^*$. The best
known result towards Conjecture~\ref{conj:op} has been obtained
by Sanders and Zhao~\cite{SaZh01}, who proved the bound
$\left\lceil\frac{5\Delta^*}{3}\right\rceil$.

Denote by $f_c(x)$ the
minimum number of colours needed to cyclically colour every plane
graph of maximum face size $x$. The value of $f_c(x)$ is known for
$x\in\{3,4\}$: $f_c(3)=4$ (the problem of finding $f_c(3)$ being
equivalent to the Four Colour Theorem proved in~\cite{ApHa89})
and $f_c(4)=6$ (see~\cite{Bor84,Bor95}). It is also known that
$f_c(5)\in\{7,8\}$ and $f_c(6)\le10$~\cite{BSZ99}, and that
$f_c(7)\le12$~\cite{Bor92}.

Conjecture~\ref{conj:kms} is trivially true
for $\l=0$, and is equivalent to the Four Colour Theorem for $\l=1$.
It is open for all other values of $\l$.
As noted in~\cite{KMS05}, if Conjecture~\ref{conj:kms} were true for $\l=2$,
it would have several interesting corollaries. Besides giving the exact
value of $f_c(5)$ (which would then be $7$), it would allow to decrease
from $16$ to $14$
(by applying a method from~\cite{KMS05}) the upper bound on the number of
colours needed to $1$-diagonally colour every plane quadrangulation
(for more details on this problem, consult~\cite{HoJe95,SaZh96,SaZh98,KMS05}).
It would also imply Wegner's conjecture on $2$-distance colourings
(i.e. colourings of squares of graphs) restricted to plane cubic graphs
since colourings of the square of a plane cubic graph are precisely its
$2$-facial colourings (refer to~\cite[Problem 2.18]{JeTo95} for more details
on Wegner's conjecture).

Let $f_f(\l)$ be the minimum number of colours needed to \l-facially
colour every plane graph. Clearly, $f_c(2\l+1)\le f_f(\l)$. So far, no
value of \l is known for which this inequality is strict. The following
problem is offered in~\cite{KMS05}.
\begin{problem}
Is it true that, for every integer $\l\ge1$, $f_c(2\l+1)=f_l(\l)$?
\end{problem}

Another conjecture that should be maybe mentioned is the so-called
$3\l$-Conjecture proposed in~\cite{DST05}, stating that every plane
triangle-free graph is \l-facially $3\l$-colourable. Similarly
as the $(3\l+1)$-Conjecture, if this conjecture were true, then
its bound would be tight
and it would have several interesting corollaries (see~\cite{DST05}
for more details).

It is proved in~\cite{KMS05} that every plane graph has an \l-facial
colouring using at most $\left\lfloor\frac{18}{5}\l\right\rfloor+2$
colours (and this bound is decreased by $1$ for $\l\in\{2,4\}$).
So, in particular, every plane graph has a $3$-facial $12$-colouring.
In this paper, we improve this last
result by proving the following theorem.
\begin{theorem}\label{theo:main}
Every plane graph is $3$-facially $11$-colourable.
\end{theorem}
%

To prove this result, we shall suppose that it is false.
In Section~\ref{sec:prop}, we will exhibit some properties
of a minimal graph (regarding the number of vertices) which
contradicts Theorem~\ref{theo:main}. Relying on these properties,
we will use the Discharging Method in Section~\ref{sec:proof}
to obtain a contradiction.


%
\section{Properties of $(3,11)$-minimal graphs}
\label{sec:prop}

Let us start this section by introducing some definitions.
A vertex of degree $d$ (respectively at least $d$, respectively at most
$d$) is said to be a {\it $d$-vertex} (respectively a {\it $(\ge
d)$-vertex}, respectively a {\it $(\le d)$-vertex}). The notion
of a {\it $d$-face} (respectively a {\it $(\le d)$-face}, respectively a {\it
$(\ge d)$-face}) is defined analogously regarding the size of a face.
An {\it \l-path} is a path of length \l.

Two faces are {\it adjacent}, or {\it neighbouring}, if they share a
common edge.
A $5$-face is {\em bad} if it is incident to at least four $3$-vertices.
It is said to be {\it very-bad} if it is incident to five
$3$-vertices.

If $u$ and $v$ are $3$-facially adjacent, then $u$ is called a
{\it $3$-facial neighbour of $v$}. The set of all $3$-facial neighbours
of $v$ is denoted by ${\mathcal N}_3(v)$. 
The {\it $3$-facial
degree} of $v$, denoted by $\deg_3(v)$, is the cardinality
of the set ${\mathcal N}_3(v)$.
A vertex is {\em dangerous} if it has degree $3$
and it is incident to a face of size three or four.
A $3$-vertex is {\em safe} if it is not dangerous,
i.e. it is not incident to a $(\le4)$-face.

Let $G=(V,E)$ be a plane graph, and ${\mathcal U}\subseteq V$. Denote
by $G_3[{\mathcal U}]$ the graph with vertex set ${\mathcal U}$ such
that $xy$ is an edge in $G_3[{\mathcal U}]$ if and only if $x$ and $y$
are $3$-facially adjacent vertices in $G$. If $c$ is a partial
colouring of $G$ and $u$ an uncoloured vertex of $G$, we denote by
$L_c(u)$ (or just $L(u)$) the set $\{x\in\{1,2,\ldots,11\}:\ \mbox{for all}\
v\in{\mathcal N}_3(u), c(v)\neq x\}$.
The graph
$G_3[{\mathcal U}]$ is {\it $L$-colourable} if there exists a
proper vertex colouring of the vertices of $G_3[{\mathcal U}]$ such that
for every $u\in{\mathcal U}$ holds $c(u)\in L(u)$.

The next two results
are used by Kr\'al', Madaras and \v{S}krekovski~\cite{KMS05}:
\begin{lemma}\label{lem:deg}
Let $v$ be a vertex whose incident faces in a plane graph $G$
are $f_1,f_2,\ldots,f_d$. Then
\[
\deg_3(v)\le\left(\sum_{i=1}^{d}\min(|f_i|,7)\right)-2d,
\]
where $|f_i|$ denotes the size of the face $f_i$.
\end{lemma}
Suppose that Theorem~\ref{theo:main} is false:
a {\it $(3,11)$-minimal} graph $G$ is a plane graph which is not
$3$-facially $11$-colourable, with $|V(G)|+|E(G)|$ as small as possible.
%
\begin{lemma}\label{lem:kms} Let $G$ be a $(3,11)$-minimal graph.
Then,
\begin{enumerate}
 \item $G$ is $2$-connected;
 \item $G$ has no separating cycle of length at most $7$;
 \item $G$ contains no  adjacent $f_1$-face and
$f_2$-face with $f_1+f_2\le9$;
 \item $G$ has no vertex whose $3$-facial degree is
less than $11$. In particular, the minimum degree of $G$
is at least three; and
 \item $G$ contains no
edge $uv$ separating two $(\ge4)$-faces with
$\deg_3(u)\le 11$ and $\deg_3(v)\le 12$.
\end{enumerate}
\end{lemma}

In the remaining of this section, we give additional local structural
properties of $(3,11)$-minimal graphs.

\begin{lemma}\label{lem:L5}
Let $G$ be a $(3,11)$-minimal graph. Suppose that
$v$ and $w$ are two adjacent $3$-vertices of $G$, both incident
to a same $5$-face and a same $6$-face. Then the size
of the third face incident to $w$ is at least $7$.
\end{lemma}

\begin{proof}
By contradiction, suppose that the size of the
last face incident to $w$ is at most $6$. Then,
according to Lemma~\ref{lem:deg}, we infer that $\deg_3(v)\le12$ and
$\deg_3(w)\le11$, but this contradicts Lemma~\ref{lem:kms}$(v)$.
\end{proof}

A {\it reducible configuration} is a (plane) graph that cannot be an
induced subgraph of a $(3,11)$-minimal graph.  The usual method to
prove that a configuration is reducible is the following: first, we
suppose that a $(3,11)$-minimal graph $G$ contains a prescribed
induced subgraph $H$. Then we contract some subgraphs $H_1,
H_2,\ldots,H_k$ of $H$. Mostly, we have $k\le2$.  This yields a proper
minor $G'$ of $G$, which by the minimality of $G$ admits a $3$-facial
$11$-colouring $c'$. The goal is to derive from $c'$ a $3$-facial
$11$-colouring $c$ of $G$, which would give a contradiction. To do so,
each non-contracted vertex $v$ of $G$ keeps its colour
$c'(v)$. Let $h_i$ be the vertex of $G'$ created by the contraction of
the vertices of $H_i$: some vertices of $H_i$ are assigned the colour
$c'(h_i)$ (in doing so, we must take care that these vertices are not
$3$-facially adjacent in $G$). Last, we show that the remaining uncoloured
vertices can also be coloured.

In other words, we show that the graph $G_3[{\mathcal U}]$ is
$L$-colourable, where for each $u\in{\mathcal U}$, $L(u)$ is the list
of the colours which are assigned to no vertex in ${\mathcal
N}_3(u)\setminus{\mathcal U}$ (defined in Section~\ref{sec:int}) and
${\mathcal U}$ is the set of uncoloured vertices. In most of the
cases, the vertices of ${\mathcal U}$ will be greedily
coloured.

In all figures of the paper, the following conventions are used: a
triangle represents a $3$-vertex, a square represents a $4$-vertex and
a circle may be any kind of vertex whose degree is at least the maximum
between three and the one it
has in the figure. The edges of each subgraph $H_i$ are drawn in bold,
and the circled vertices are the vertices of ${\mathcal U}=\{u_1, u_2,
\dots \}$.  A dashed edge between two vertices indicates a path of
length at least one between those two vertices.  An (in)equality
written in a bounded region denotes a face whose size achieves the
(in)equality. Last, vertices which are assigned the colour $c'(h_i)$
are denoted by $v$, $w$, $t$ if a unique subgraph is contracted or
by $x_1,x_2$ for $i=1$ and $y_1,y_2$ for $i=2$ if two
subgraphs are contracted.
%
%
%
%

\begin{lemma}
Configurations in Figures~\ref{fig:c1},~\ref{fig:c2} and~\ref{fig:c3} are reducible.
\end{lemma}
\begin{proof}
Let $H$ be an induced subgraph of $G$. We shall suppose that $H$ is
isomorphic to one of the configurations stated and derive a way
to construct a
$3$-facial $11$-colouring of $G$, a contradiction.

\begin{figure}[p]
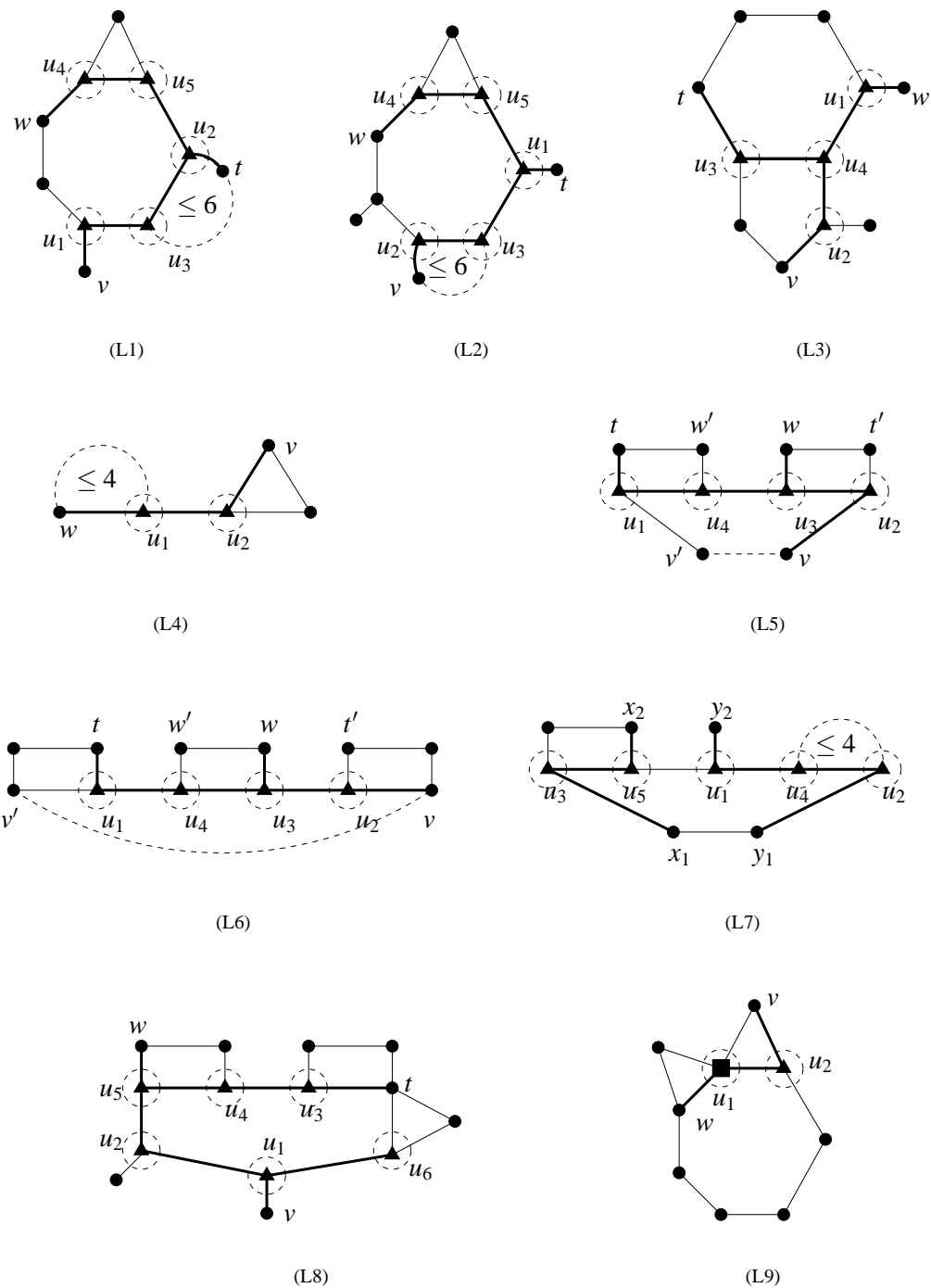

\centering
  \subfigure[(L1)]{\input{L2a.pstex_t}}
\hfill
  \subfigure[(L2)]{\input{L2b.pstex_t}}
\hfill
  \subfigure[(L3)]{\input{L3.pstex_t}}\\
  \subfigure[(L4)]{\input{L4.pstex_t}}
  \hfill
  \subfigure[(L5)]{\input{L9a.pstex_t}}\\
  \subfigure[(L6)]{\input{L9b.pstex_t}}
  \hfill
  \subfigure[(L7)]{\input{L10.pstex_t}}\\
  \subfigure[(L8)]{\input{L11.pstex_t}}
  \hfil
  \subfigure[(L9)]{\label{fig:L29}\input{L29.pstex_t}}\\
 \caption{Reducible configurations (L1)--(L9).}
 \label{fig:c1}
\end{figure}

\noindent
\paragraph{L1.}

\noindent
Suppose that $H$ is
isomorphic to the configuration (L1) of Figure~\ref{fig:c1}.
Denote by $H_1$ the subgraph induced by the bold edges.
Contract the vertices of $H_1$, thereby creating a new vertex $h_1$.
By minimality of $G$, let $c'$ be a
$3$-facial $11$-colouring of the obtained graph.
Assign to each vertex $x$ not in $H_1$ the colour $c'(x)$,
and to each of $v,w,t$ the colour $c'(h_1)$. Observe that
no two vertices among $v,w,t$ are $3$-facially adjacent in $G$, otherwise
there would be a $(\le7)$-separating cycle in $G$, thereby contradicting
Lemma~\ref{lem:kms}$(ii)$.
According to Lemma~\ref{lem:deg}, $\deg_3(u_1)\le15$,
$\deg_3(u_i)\le14$ if $i\in\{2,3\}$ and
$\deg_3(u_i)\le11$ if $i\in\{4,5\}$.
Note that any two vertices of ${\mathcal U}=\{u_1,u_2,\ldots,u_5\}$ are $3$-facially
adjacent, that is $G_3[{\mathcal U}]\simeq K_5$.
Hence, the number of coloured $3$-facial neighbours of $u_1$ is at most
$11$, i.e. $|{\mathcal N}_3(u_1)\setminus\{u_2,u_3,u_4,u_5\}|\ge11$.
Moreover, at least two of them are assigned the same
colour, namely $v$ and $w$.
Therefore, $|L(u_1)|\ge1$.
For $i\in\{2,3\}$, the vertex $u_i$ has at most $10$ coloured $3$-facial
neighbours. Furthermore, at least two $3$-facial neighbours of $u_2$
are identically coloured, namely $w$ and $t$. Thus, $|L(u_2)|\ge2$.
Now, observe that at least three $3$-facial neighbours of $u_3$ are
coloured the same, namely $v,w$ and $t$. Hence, $|L(u_3)|\ge3$.
For $i\in\{4,5\}$, the vertex $u_i$ has at most $7$ coloured
$3$-facial neighbours. Thus, $|L(u_4)|\ge4$, and because at least two
$3$-facial neighbours of $u_5$ are identically coloured ($w$ and $t$),
$|L(u_5)|\ge5$. So, the graph $G_3[{\mathcal U}]$ is greedily
$L$-colourable, according to the ordering $u_1,u_2,u_3,u_4,u_5$. This
allows us to extend $c$ to a $3$-facial $11$-colouring of $G$.

\noindent
\paragraph{L2.}

\noindent
Suppose that $H$ is isomorphic to the configuration (L2)
of Figure~\ref{fig:c1}.
Let $c'$ be a $3$-facial $11$-colouring of the minor
of $G$ obtained by contracting the bold edges into a single vertex $h_1$.
Let $c(x)=c'(x)$ for every vertex $x\neq h_1$. Define
$c(v)=c(w)=c(t)=c'(h_1)$. The obtained colouring is still $3$-facial
since no two vertices among $v,w,t$ are $3$-facially adjacent in $G$ by
Lemma~\ref{lem:kms}$(ii)$. Note that $G_3[{\mathcal U}]\simeq K_5$.
In particular, each vertex $u_i$ has four uncoloured $3$-facial
neighbours.
By Lemma~\ref{lem:deg}, $\deg_3(u_1)\le15,
\deg_3(u_i)\le14$ if $i\in\{2,3\}$ and $\deg_3(u_i)\le11$ if $i\in\{4,5\}$.
Moreover, each of $u_1$ and $u_2$ has
at least two $3$-facial neighbours coloured the same; for $u_1$, these
vertices are $w,t$ and for $u_2$ they are $w,v$. So, there exists at least one colour which is
assigned to no vertex of ${\mathcal N}_3(u_1)$ and at least two
colours assigned to no vertex of ${\mathcal N}_3(u_2)$.
Also, $u_3$ has at least three $3$-facial neighbours coloured the
same, namely $w$, $v$ and $t$, hence
at least three colours are assigned to no vertex of ${\mathcal N}_3(u_3)$.
Therefore, $|L(u_1)|\ge1, |L(u_2)|\ge2$ and $|L(u_3)|\ge3$.
Furthermore, $|L(u_4)|\ge4$ and $|L(u_5)|\ge5$ because $w$ and $t$ are both
$3$-facial neighbours of $u_5$. So $G_3[{\mathcal U}]$ is
$L$-colourable, and hence $G$ is $3$-facially $11$-colourable.

%
\noindent
\paragraph{L3.}

\noindent
Suppose that $H$ is isomorphic to the configuration (L3) of
Figure~\ref{fig:c1}.  Contract the bold edges into a new vertex $h_1$,
and let $c'$ be a $3$-facial $11$-colouring of the obtained
graph. This colouring can be extended to a $3$-facial $11$-colouring
$c$ of $G$ as follows: first, let $c(v)=c(w)=c(t)=c'(h_1)$. Note that
no two of these vertices can be $3$-facially adjacent in $G$ without
contradicting Lemma~\ref{lem:kms}$(ii)$. By Lemma~\ref{lem:deg},
$\deg_3(u_1)\le14$, $\deg_3(u_2)\le13$ and for $i\in\{3,4\}$,
$\deg_3(u_i)\le12$. Observe that $G_3[{\mathcal U}]\simeq
K_4$. Moreover, each of $u_1,u_2,u_3$ has a set of two $3$-facial
neighbours coloured by $c'(h_1)$. These sets are $\{w,t\}$, $\{w,v\}$ and
$\{v,t\}$ for $u_1,u_2$ and $u_3$, respectively.
Thus, $|L(u_1)|\ge1$, $|L(u_2)|\ge2$ and
$|L(u_3)|\ge3$. Also $|L(u_4)|\ge4$ because $u_4$ has at least three
identically coloured $3$-facial neighbours, namely $v,w$ and $t$.  Hence,
$G_3[{\mathcal U}]$ is $L$-colourable, so $G$ is $3$-facially
$11$-colourable.

\noindent
\paragraph{L4.}

\noindent
Let $c'$ be a $3$-facial $11$-colouring
of the graph obtained by
contracting the bold edges into a new vertex $h_1$.
Define $c(x)=c'(x)$ if $x\notin\{v,w,u_1,u_2\}$ and
$c(v)=c(w)=c'(h_1)$. Observe that $v$ and $w$ cannot
be $3$-facially adjacent in $G$ since $G$ has no
small separating cycle according to Lemma~\ref{lem:kms}$(ii)$.
By Lemma~\ref{lem:deg}, $\deg_3(u_1)\le12$ and
$\deg_3(u_2)\le11$. Furthermore, both $u_1$ and $u_2$
have two $3$-facial neighbours identically coloured, namely
$v$ and $w$. Moreover, $u_1$ and $u_2$ are $3$-facially adjacent,
hence $|L(u_1)|\ge1$ and $|L(u_2)|\ge2$. Therefore, $c$ can be extended
to a $3$-facial $11$-colouring of $G$.

%
%
%
%
%

%
\noindent
\paragraph{L5.}

\noindent
First, observe that since $G$ is a plane graph,
if $v\in{\mathcal N}_3(t)$ then $v'\notin{\mathcal N}_3(t')$.
So, by symmetry, we may assume that $v$ and $t$
are not $3$-facially adjacent in $G$. Now, contract the bold edges
into a new vertex $h_1$. Again, denote by $c'$ a $3$-facial
$11$-colouring of the obtained graph, and define $c$ to be equal
to $c'$ on all vertices of
$V(G)\setminus\{v,w,t,u_1,u_2,u_3,u_4\}$.
Let $c(v)=c(w)=c(t)=c'(h_1)$.
Note that the partial colouring
$c$ is still $3$-facial due to the above assumption.
The graph $G_3[{\mathcal U}]$ is isomorphic to $K_4$, and according
to Lemma~\ref{lem:deg}, $\deg_3(u_i)\le12$ for all $i\in\{1,2,3,4\}$.
Moreover, for $i\in\{2,3\}$, the vertex $u_i$ has at least two $3$-facial
neigbhours that are coloured the same, namely
$v$ and $w$. Last, the vertex $u_4$ has at least three
such $3$-facial neighbours, namely $v$, $w$, $t$. Therefore,
$|L(u_1)|\ge2$, $|L(u_i)|\ge3$ for $i\in\{2,3\}$ and $|L(u_4)|\ge4$.
So, $G_3[{\mathcal U}]$ is $L$-colourable, and hence $G$ is $3$-facially
$11$-colourable.

%
\noindent
\paragraph{L6.}

\noindent
The same remark as in the previous configuration allows us to assume that
$t\notin{\mathcal N}_3(v)$. Again, the graph obtained by
contracting the bold edges into a new vertex $h_1$ admits a $3$-facial
$11$-colouring $c'$. As before, define a $3$-facial
$11$-colouring $c$ of the graph induced by $V(G)\setminus{\mathcal
U}$.  Then, for every $i\in\{1,2,3,4\},
\deg_3(u_i)\le12$ and $G_3[{\mathcal U}]\simeq K_4$. Thus, $|L(u_1)|\ge2$
and $|L(u_2)|\ge2$. Remark that $u_3$ has at least two identically
coloured $3$-facial neighbours, namely $v$ and $w$, so
$|L(u_3)|\ge3$. Last, the vertex $u_4$ has at least three such
neighbours, hence $|L(u_4)|\ge4$. Therefore, the graph $G_3[{\mathcal U}]$
is $L$-colourable, and so the graph $G$ admits a $3$-facial
$11$-colouring.

\noindent
\paragraph{L7.}

\noindent
Let $H_1$ be the path $x_1u_3u_5x_2$, $H_2$ the path
$y_1u_2u_4u_1y_2$ and $c'$ a $3$-facial colouring of the graph obtained
from $G$ by contracting each path $H_i$ into a vertex $h_i$.
Notice that $c'(h_1)\neq c'(h_2)$.
For every $v\notin V(H_1)\cup V(H_2)$, let
$c(v)=c'(v)$. Observe that $x_1$ and $x_2$ cannot
be $3$-facially adjacent in $G$, otherwise $G$ would have
a separating $(\le7)$-cycle, contradicting Lemma~\ref{lem:kms}$(iii)$.
Note that the same holds for $y_1$ and $y_2$; therefore defining
$c(x_1)=c(x_2)=c'(h_1)$ and
$c(y_1)=c(y_2)=c'(h_2)$ yields a partial
$3$-facial $11$-colouring of $G$, since $c'(h_1)\neq c'(h_2)$.
It remains to colour the vertices of ${\mathcal
U}=\{u_1,u_2,\ldots,u_5\}$. Note that $G_3[{\mathcal U}]\simeq K_5$.
According to Lemma~\ref{lem:kms}$(ii)$,
$\deg_3(u_1)\le15$ and $\deg_3(u_i)\le12$ if $i\ge2$.
The number of coloured $3$-facial neighbours of $u_1$, i.e. its number
of $3$-facial neighbours in $V(G)\setminus\{u_2,u_3,u_4,u_5\}$, is at most
$11$ because each $u_i$ with $i\ge2$ is a $3$-facial neighbour of
$u_1$. Furthermore, $u_1$ has two $3$-facial neighbours coloured with
the same colour, namely $x_1$ and $x_2$. Hence, $|L(u_1)|\ge1$.
The vertex $u_2$ has four uncoloured $3$-facial neighbours, so $|L(u_2)|\ge3$.
For $i\in\{3,4\}$, the vertex $u_i$ has at least two
$3$-facial neighbours coloured
the same, namely $x_1,x_2$ for $u_3$, and $y_1,y_2$ for
$u_4$, so $|L(u_i)|\ge4$. Finally, observe that $u_5$ has two
pairs of identically coloured $3$-facial neighbours; the first pair
being $x_1,x_2$ and the second $y_1,y_2$.
Thus, $|L(u_5)|\ge5$, hence the graph $G_3[{\mathcal U}]$ is
$L$-colourable, which yields a contradiction.

%
%
%
%
%
%
\begin{figure}[p]
\centering
  \subfigure[(L10)]{\label{fig:L12}\input{L12bad.pstex_t}}
  \hfil
  \subfigure[(L11)]{\label{fig:L12b}\input{L12b.pstex_t}}\\
  \subfigure[(L12)]{\label{fig:L13}\input{L13.pstex_t}}
\hfill
  \subfigure[(L13)]{\label{fig:L14}\input{L14.pstex_t}}
\hfill
  \subfigure[(L14)]{\label{fig:L15}\input{L15.pstex_t}}
\hfill
  \subfigure[(L15)]{\label{fig:L16}\input{L16.pstex_t}}
\hfill
  \subfigure[(L16)]{\label{fig:L17}\input{L17.pstex_t}}\\

 \caption{Reducible configurations (L10)--(L16).}
 \label{fig:c2}
\end{figure}

\noindent
\paragraph{L8.}

\noindent
We contract the bold edges into a new vertex $h_1$,
take a $3$-facial $11$-colouring of the graph obtained, and
define a $3$-facial $11$-colouring $c$ of $V(G)\setminus{\mathcal U}$
as usual. By Lemma~\ref{lem:deg}, $\deg_3(u_i)\le15$ if $i\in\{1,2\}$,
$\deg_3(u_i)\le12$ if $i\in\{3,4,5\}$ and $\deg_3(u_6)\le11$.
Moreover, $G_3[{\mathcal U}]\simeq K_6$. As $v,w$ and $t$ are coloured the
same, and $\{v,w\}\subset{\mathcal N}_3(u_i)$ for $i\in\{2,5\}$,
$\{w,t\}\subset{\mathcal N}_3(u_4)$ and $\{v,t\}\subset{\mathcal
N}_3(u_5)$,
we obtain $|L(u_i)|\ge i$ for every $i\in\{1,2,3,4,5,6\}$. Thus, the
graph $G_3[{\mathcal U}]$ is $L$-colourable, and hence
$G$ admits a $3$-facial $11$-colouring.

\noindent
\paragraph{L9.}

\noindent
We contract the bold edges into a new vertex,
take a $3$-facial $11$-colouring of the graph obtained, and
define a $3$-facial $11$-colouring of $V(G)\setminus{\mathcal U}$
as usual.
Then, $G_3[{\mathcal U}]\simeq K_2$. Moreover,
$\deg_3(u_1)\le12$ and
$\deg_3(u_2)\le11$.
Furthermore,
$\{v,w\}\subset{\mathcal N}_3(u_i)$ for $i\in\{1,2\}$.
Thus, we infer $|L(u_i)|\ge i$ for $i\in\{1,2\}$.
Therefore, $G_3[{\mathcal U}]$ is $L$-colourable.

%
%
%
%

\noindent
\paragraph{L10.}

\noindent
We contract the bold edges into a new vertex,
take a $3$-facial $11$-colouring of the graph obtained, and
define a $3$-facial $11$-colouring of $V(G)\setminus{\mathcal U}$
as usual.
Then, $G_3[{\mathcal U}]\simeq K_4$. Moreover,
$\deg_3(u_1)\le13$, $\deg_3(u_2)\le12$ and
$\deg_3(u_i)\le11$ for $i\in\{3,4\}$.
Furthermore,
$\{v,w\}\subset{\mathcal N}_3(u_i)$ for $i\in\{1,4\}$.
Thus, we infer $|L(u_i)|\ge 2$ for $i\in\{1,2\}$, and
$|L(u_i)|\ge i$ for $i\in\{3,4\}$.
Therefore, $G_3[{\mathcal U}]$ is $L$-colourable.

%

\noindent
\paragraph{L11.}

\noindent
We contract the bold edges into a new vertex $h_1$,
take a $3$-facial $11$-colouring of the graph obtained, and
define a $3$-facial $11$-colouring $c$ of $V(G)\setminus{\mathcal U}$
as usual. By Lemma~\ref{lem:deg}, $\deg_3(u_1)\le15$ and
$\deg_3(u_i)\le11$ if $i\in\{2,3,4,5\}$.
Moreover, $G_3[{\mathcal U}]\simeq K_5$. As $v$ and $w$ are coloured the
same, and $\{v,w\}\subset{\mathcal N}_3(u_i)$ for $i\in\{1,4,5\}$,
we obtain $|L(u_1)|\ge 1$, $|L(u_i)|\ge4$ if $i\in\{2,3\}$
and $|L(u_i)|\ge5$ if $i\in\{4,5\}$. Thus, the
graph $G_3[{\mathcal U}]$ is $L$-colourable, and hence
$G$ admits a $3$-facial $11$-colouring.

\noindent
\paragraph{L12.}

\noindent
Let $c'$ be a $3$-facial $11$-colouring of the graph $G'$
obtained by contracting the bold edges into a new vertex $h_1$.
Define $c(x)=c'(x)$
for every vertex $x\in V(G)\cap V(G')$, and let $c(v)=c(w)=c'(h_1)$.
By Lemma~\ref{lem:deg}, $\deg_3(u_i)\le15$ for $i\in\{1,2\}$ and
$\deg_3(u_i)\le11$ for $i\in\{3,4,5\}$.
Moreover, $G_3[{\mathcal U}]\simeq K_6$.
Hence, $|L(u_1)|\ge1$ and $|L(u_i)|\ge i$ for $i\in\{3,4,5\}$.
As $v$ and $w$ are coloured the
same, and $\{v,w\}\subset{\mathcal N}_3(u_i)$ for $i\in\{2,6\}$,
we infer that $|L(u_2)|\ge2$ and $|L(u_6)|\ge6$.
Thus, the graph $G$ is $3$-facially $11$-colourable.

\noindent
\paragraph{L13.}

\noindent
Let us define the partial $3$-facial $11$-colouring $c$ as
always, regarding the bold edges and the vertices $v$ and $w$.
From Lemma~\ref{lem:deg} we get $\deg_3(u_1)\le15$,
$\deg_3(u_i)\le12$ for $i\in\{2,3,4\}$ and $\deg_3(u_5)\le11$.
Moreover, since $G_3[{\mathcal U}]\simeq K_5$ and
$\{v,w\}\subset{\mathcal N}_3(u_i)$ for $i\in\{1,4,5\}$,
we obtain $|L(u_1)|\ge1$, $|L(u_i)|\ge3$ for $i\in\{2,3\}$,
$|L(u_4)|\ge4$ and $|L(u_5)|\ge5$. Therefore, $G_3[{\mathcal U}]$
is $L$-colourable.

\noindent
\paragraph{L14.}

\noindent
Define the partial $3$-facial $11$-colouring $c$ as
usual, regarding the bold edges and the vertices $v$ and $w$.
By Lemma~\ref{lem:deg}, $\deg_3(u_1)\le15$ and
$\deg_3(u_i)\le11$ for $i\in\{2,3,4,5\}$.
Moreover, since $G_3[{\mathcal U}]\simeq K_5$ and
$\{v,w\}\subset{\mathcal N}_3(u_i)$ for $i\in\{1,5\}$,
we obtain $|L(u_1)|\ge1$, $|L(u_i)|\ge4$ for $i\in\{2,3,4\}$ and
$|L(u_5)|\ge5$. Therefore, $G_3[{\mathcal U}]$
is $L$-colourable.

\noindent
\paragraph{L15.}

\noindent
Let us define the partial $3$-facial $11$-colouring $c$ as
always, regarding the bold edges and the vertices $v$ and $w$.
Again, $G_3[{\mathcal U}]\simeq K_5$.
From Lemma~\ref{lem:deg} we get $\deg_3(u_1)\le15$ and
$\deg_3(u_i)\le11$ if $i\in\{2,3,4,5\}$.
Moreover, since
$\{v,w\}\subset{\mathcal N}_3(u_i)$ for $i\in\{1,5\}$,
we obtain $|L(u_1)|\ge1$, $|L(u_i)|\ge4$ for $i\in\{2,3,4\}$ and
$|L(u_5)|\ge5$. Therefore, $G_3[{\mathcal U}]$
is $L$-colourable.

\noindent
\paragraph{L16.}

\noindent
Define the partial $3$-facial $11$-colouring $c$ as
always, regarding the bold edges and the vertices $v,w$ and $t$.
Then, $G_3[{\mathcal U}]\simeq K_5$ and
$\deg_3(u_i)\le15$ for $i\in\{1,2\}$,
$\deg_3(u_i)\le12$ for $i\in\{3,4\}$ and
$\deg_3(u_5)\le11$.
Moreover, notice that
$\{v,t\}\subset{\mathcal N}_3(u_i)$ for $i\in\{1,4\}$,
$\{v,w,t\}\subset{\mathcal N}_3(u_2)$ and
$\{v,w\}\subset{\mathcal N}_3(u_5)$.
Thus, we obtain $|L(u_1)|\ge1$, $|L(u_2)|\ge2$, $|L(u_3)|\ge3$,
$|L(u_4)|\ge4$ and
$|L(u_5)|\ge5$. Therefore, $G_3[{\mathcal U}]$
is $L$-colourable.

\noindent
\paragraph{L17.}

\noindent
Define the partial $3$-facial $11$-colouring $c$ as
always, regarding the bold edges and the vertices $v,w$ and $t$.
Then, $G_3[{\mathcal U}]\simeq K_5$ and
$\deg_3(u_i)\le15$ for $i\in\{1,2\}$,
$\deg_3(u_3)\le12$ and
$\deg_3(u_i)\le11$ for $i\in\{4,5\}$.
Moreover, notice that
$\{v,t\}\subset{\mathcal N}_3(u_i)$ for $i\in\{1,5\}$,
$\{v,w,t\}\subset{\mathcal N}_3(u_2)$ and
$\{v,w\}\subset{\mathcal N}_3(u_3)$.
Thus, we obtain $|L(u_1)|\ge1$, $|L(u_2)|\ge2$, $|L(u_i)|\ge4$
for $i\in\{3,4\}$ and
$|L(u_5)|\ge5$. Therefore, $G_3[{\mathcal U}]$
is $L$-colourable.

\noindent
\paragraph{L18.}

\noindent
Let us define the partial $3$-facial $11$-colouring $c$ as
always, regarding the bold edges and the vertices $v$ and $w$.
Then, $G_3[{\mathcal U}]\simeq K_3$,
$\deg_3(u_1)\le13$ and
$\deg_3(u_i)\le11$ for $i\in\{2,3\}$.
Moreover,
$\{v,w\}\subset{\mathcal N}_3(u_i)$ for $i\in\{1,2,3\}$.
Thus, we obtain $|L(u_1)|\ge1$ and $|L(u_i)|\ge3$
for $i\in\{2,3\}$.
Therefore, $G_3[{\mathcal U}]$ is $L$-colourable.

\noindent
\paragraph{L19.}

\noindent
Again, $G_3[{\mathcal U}]\simeq K_5$ and
$\deg_3(u_i)\le15$ for $i\in\{1,2\}$ while
$\deg_3(u_i)\le11$ for $i\in\{3,4,5\}$.
Furthermore,
$\{v,w\}\subset{\mathcal N}_3(u_i)$ for $i\in\{1,3,4\}$,
$\{v,t\}\subset{\mathcal N}_3(u_5)$ and
$\{v,w,t\}\subset{\mathcal N}_3(u_2)$.
Thus, we deduce $|L(u_1)|\ge1$, $|L(u_2)|\ge2$ and $|L(u_i)|\ge5$
for $i\in\{3,4,5\}$.
Therefore, $G_3[{\mathcal U}]$ is $L$-colourable.

\begin{figure}[tp]
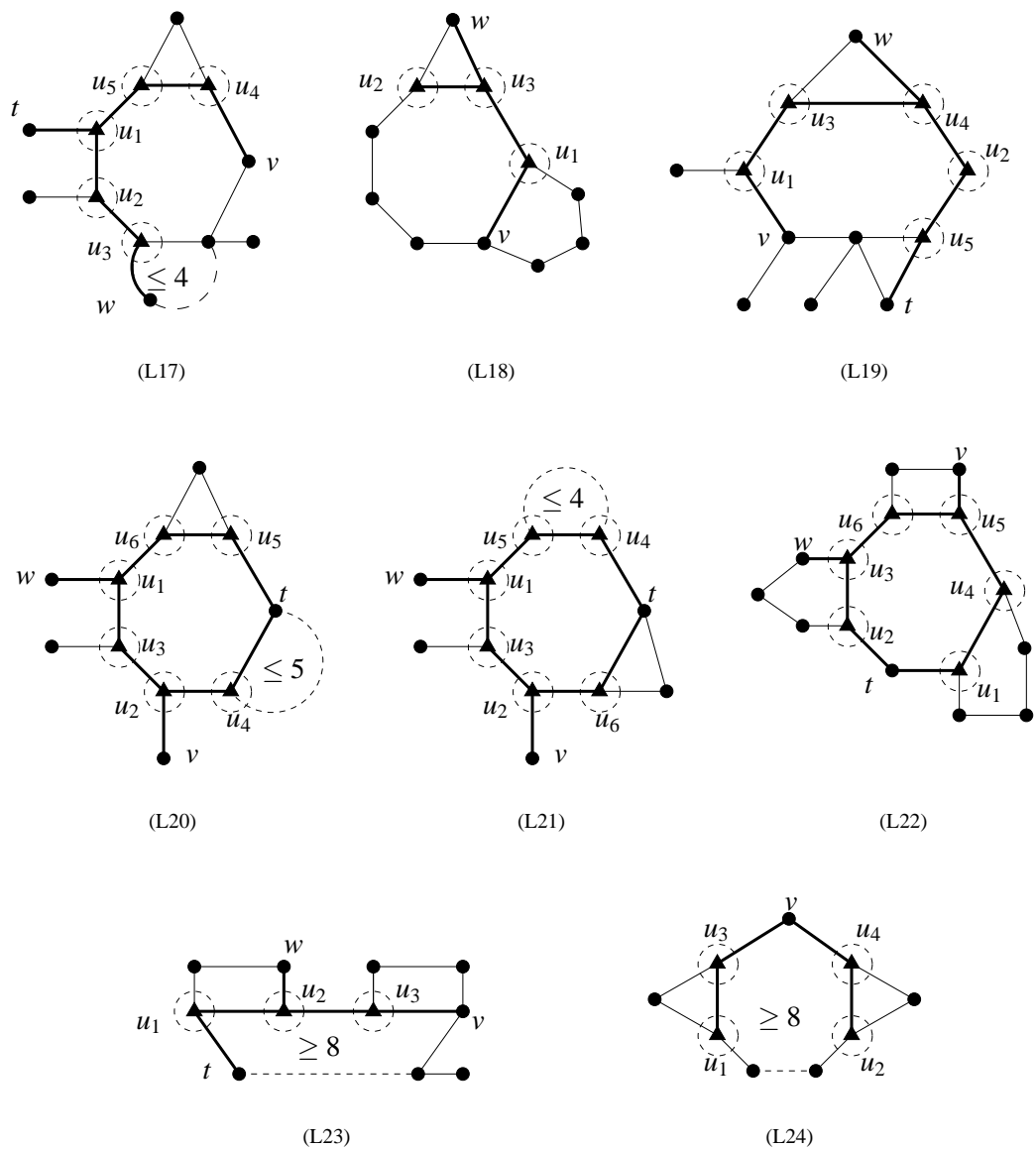

\centering
  \subfigure[(L17)]{\label{fig:L17b}\input{L17b.pstex_t}}
\hfill
  \subfigure[(L18)]{\label{fig:L18}\input{L18.pstex_t}}
\hfill
  \subfigure[(L19)]{\label{fig:L19}\input{L19.pstex_t}}\\
  \subfigure[(L20)]{\label{fig:L20}\input{L20.pstex_t}}
\hfill
  \subfigure[(L21)]{\label{fig:L20b}\input{L20b.pstex_t}}
\hfill
  \subfigure[(L22)]{\label{fig:L21}\input{L21p.pstex_t}}\\
  \subfigure[(L23)]{\label{fig:L30}\input{L30.pstex_t}}
\hfil
  \subfigure[(L24)]{\label{fig:L31}\input{L31.pstex_t}}
 \caption{Reducible configurations (L17)--(L24).}
 \label{fig:c3}
\end{figure}

\noindent
\paragraph{L20.}

\noindent
Here,
$G_3[{\mathcal U}]\simeq K_6$. Also,
$\deg_3(u_i)\le15$ for $i\in\{1,2,3\}$,
$\deg_3(u_4)\le13$ and
$\deg_3(u_i)\le11$ for $i\in\{5,6\}$.
Furthermore,
$\{w,t\}\subset{\mathcal N}_3(u_i)$ for $i\in\{1,6\}$,
$\{v,w,t\}\subset{\mathcal N}_3(u_3)$ and
$\{v,t\}\subset{\mathcal N}_3(u_i)$ for $i\in\{2,4\}$.
Thus, we infer $|L(u_i)|\ge2$ for $i\in\{1,2\}$, $|L(u_3)|\ge3$,
$|L(u_4)|\ge4$, $|L(u_5)|\ge5$
and $|L(u_6)|\ge6$.
Therefore, $G_3[{\mathcal U}]$ is $L$-colourable.

\noindent
\paragraph{L21.}

\noindent
Again
$G_3[{\mathcal U}]\simeq K_6$. Also,
$\deg_3(u_i)\le15$ for $i\in\{1,2,3\}$,
$\deg_3(u_i)\le12$ for $i\in\{4,5\}$ and
$\deg_3(u_6)\le11$.
Furthermore,
$\{w,t\}\subset{\mathcal N}_3(u_i)$ for $i\in\{1,5\}$,
$\{v,w,t\}\subset{\mathcal N}_3(u_3)$ and
$\{v,t\}\subset{\mathcal N}_3(u_i)$ for $i\in\{2,6\}$.
Thus, we infer $|L(u_i)|\ge2$ for $i\in\{1,2\}$ and $|L(u_i)|\ge i$
for $i\in\{3,4,5,6\}$.
Therefore, $G_3[{\mathcal U}]$ is $L$-colourable.

\noindent
\paragraph{L22.}

\noindent
In this case,
$G_3[{\mathcal U}]\simeq K_6$. Also,
$\deg_3(u_i)\le13$ for $i\in\{1,2,3,4\}$ and
$\deg_3(u_i)\le12$ for $i\in\{5,6\}$.
Furthermore,
$\{v,t\}\subset{\mathcal N}_3(u_i)$ for $i\in\{4,5\}$,
$\{v,w,t\}\subset{\mathcal N}_3(u_6)$ and
$\{w,t\}\subset{\mathcal N}_3(u_i)$ for $i\in\{2,3\}$.
Thus, we infer $|L(u_1)|\ge3$, $|L(u_i)|\ge4$ for $i\in\{2,3,4\}$,
$|L(u_5)|\ge5$ and $|L(u_6)|\ge6$.
Therefore, $G_3[{\mathcal U}]$ is $L$-colourable.

%
%
%
%
%
%
%
%

%
\noindent
\paragraph{L23.}

\noindent
In this case,
$G_3[{\mathcal U}]\simeq K_3$. Also,
$\deg_3(u_i)\le12$ for $i\in\{1,2,3\}$.
Moreover,
$\{v,w,t\}\subset{\mathcal N}_3(u_i)$ for $i\in\{1,2,3\}$.
Thus, we infer $|L(u_i)|\ge3$ for
$i\in\{1,2,3\}$.
Therefore, $G_3[{\mathcal U}]$ is $L$-colourable.

\noindent
\paragraph{L24.}

\noindent
Define the partial colouring $c$ as always, regarding the bold edges
and the vertex $v$. Remark that $G_3[{\mathcal U}]$ is isomorphic
to the complete graph on four vertices minus one edge $K_{4}^{-}$,
since $u_1\notin
{\mathcal N}_3(u_2)$ (because the face has size at least $8$).
By Lemma~\ref{lem:deg}, $\deg_3(u_i)\le11$ for every $i\in\{1,2,3,4\}$.
Thus, $|L(u_i)|\ge2$ for $i\in\{1,2\}$ and $|L(u_i)|\ge3$ for
$i\in\{3,4\}$. Hence, the graph $G_3[{\mathcal U}]$ is $L$-colourable.
This assertion can be directly checked, or seen as a consequence of a
theorem independently proved by Borodin~\cite{Bor77} and
Erd\H{o}s, Rubin and Taylor~\cite{ERT79} (see also~\cite{Tho97}),
stating that a
connected graph is degree-choosable unless it is a {\it Gallai tree}, 
that is each of its blocks
is either complete or an odd cycle.
%
\end{proof}

\begin{corollary}\label{cor}
Every $(3,11)$-minimal graph $G$ has the following properties:
\begin{enumerate}
\item Let $f_1,f_2$ be two $5$-faces of $G$ with a common edge
$xy$. Then, $x$ and $y$ are not both $3$-vertices.
\item Let $f$ be a $7$-face whose every incident vertex is a
$3$-vertex. If $f$ is adjacent to a $3$-face, then every other 
face adjacent to $f$
is a $(\ge 7)$-face.
\item If two adjacent dangerous vertices do not lie on a same
$(\le4)$-face, then none of them is incident to a $3$-face.
\item Two dangerous vertices incident to a same $6$-face
are not adjacent.
\item There cannot be four consecutive dangerous vertices
incident to a same $(\ge6)$-face.
\item A very-bad face is adjacent to at least three $(\ge7)$-faces.
\item A bad face is adjacent to at least two $(\ge7)$-faces.
\end{enumerate}
\end{corollary}

\begin{proof}~
\begin{enumerate}

\item By Lemma~\ref{lem:kms}$(v)$, $\deg_3(x)+\deg_3(y)\ge23$.
By Lemma~\ref{lem:deg}, the $3$-facial degree of a $3$-vertex incident
to two $5$-faces is at most $11$. Hence at least one of
$x$ and $y$ is a $(\ge4)$-vertex.

\item First note that, according to Lemma~\ref{lem:kms}$(iii)$, the faces
adjacent to both $f$ and the $3$-face has size at least $7$. Hence, $f$ is
adjacent to at most four $(\le6)$-faces. Now, the assertion directly
follows from the reducibility of the configurations (L1) and (L2) of
Figure~\ref{fig:c1}.

\item This follows from the reducibility of the configuration (L4) of
Figure~\ref{fig:c1}.

\item Suppose the contrary, and let $x$ and $y$ be two such vertices.
By Lemma~\ref{lem:kms}$(iii)$, a $6$-face is not adjacent to a $3$-face,
hence both $x$ and $y$ are incident to a $4$-face. Then,
$\deg_3(x)\le11$ and $\deg_3(y)\le11$, which contradicts
Lemma~\ref{lem:kms}$(v)$.

\item Suppose that the assertion is false. Then, according to the third item of this corollary,
the graph $G$ must contain the configuration (L5) or (L6) of
Figure~\ref{fig:c1}, which are both reducible.
\item Let $f$ be a very-bad face. By the first item of this corollary and Lemma~\ref{lem:L5},
two adjacent $(\le6)$-faces
cannot be both adjacent to $f$. Hence, $f$ is adjacent to at most two
such faces.
\item Let $f$ be a bad face, and denote by $\alpha_i$, $i\in\{1,2,3,4,5\}$
its incident vertices in clockwise order. Without loss of generality,
assume that, for every $i\in\{1,2,3,4\}$, $\alpha_i$ is a dangerous vertex.
For $i\in\{1,2,3,4\}$, denote by $f_i$ the face adjacent to $f$
and incident to both $\alpha_i$ and $\alpha_{i+1}$. According to the first item of
this corollary and Lemma~\ref{lem:L5}, at most two faces among $f_1,f_2,f_3,f_4$ can
be $(\le6)$-faces. This concludes the proof.
\end{enumerate}
\end{proof}

\section{Proof of Theorem~\ref{theo:main}}
\label{sec:proof}

Suppose that Theorem~\ref{theo:main} is false, and let $G$ be
a $(3,11)$-minimal graph. We shall get a contradiction by using the Discharging
Method. Here is an overview of the proof:
each vertex and face is assigned an initial charge.
The total sum of the charges is known to be negative by Euler's Formula.
Then, some redistribution rules are defined, and
each vertex and face gives or receives some charge according
to these rules. The total sum of the charges is not changed
during this step, but at the end we shall show, by case analysis,
that the charge of each vertex and each face is non-negative, a contradiction.

\paragraph{Initial charge.} First, we assign a charge to each vertex and face.
For every $v\in V(G)$, we define the initial charge
\[ \ch(v) = d(v) - 4, \]
where $d(v)$ is the degree of the vertex $v$ in $G$.
Similarly, for every $f\in F(G)$, where $F(G)$ is the set of faces of $G$,
we define the initial charge
\[ \ch(f) = r(f) -4, \]
with $r(f)$ the length of the face $f$.
By Euler's formula the total sum is
\[ \sum_{v\in V(G)} \ch(v) + \sum_{f\in F(G)} \ch(f) = -8. \]

\paragraph{Rules.} We use the following discharging rules to
redistribute the initial charge.\\

\noindent
{\bf Rule R1.} {\it A $(\ge 5)$-face sends $1/3$ to each of its incident safe
vertices and $1/2$ to each of its incident dangerous vertices.}
\medskip

\bigskip
\noindent
{\bf Rule R2.} {\it A $(\ge7)$-face sends $1/3$ to each adjacent
$3$-face.}
\medskip

\bigskip
\noindent
{\bf Rule R3.} {\it A $(\ge7)$-face sends $1/6$ to each adjacent
bad face.}
\medskip

\bigskip
\noindent
{\bf Rule R4.} {\it A $6$-face sends $1/12$ to each adjacent
very-bad face.}
\medskip

\bigskip
\noindent
{\bf Rule R5.} {\it A $(\ge5)$-vertex $v$ gives $2/3$ to an incident
face $f$ if and only if there exist two $3$-faces both incident to $v$
and both adjacent to $f$. (Note that the size of such a face $f$ is at
least $7$.)}
\medskip

We shall prove now that the final charge $\ch^*(x)$ of every
$x\in V(G)\cup F(G)$ is non-negative. Therefore, we obtain
\[
-8=\sum_{v\in V(G)}\ch(v)+\sum_{f\in F(G)}\ch(f)=\sum_{v\in
V(G)}\ch^*(v)+\sum_{f\in F(G)}\ch^*(f)\ge0,
\]
a contradiction.

\paragraph{Final charge of vertices.}

First, as noticed in Lemma~\ref{lem:kms}$(iv)$, $G$ has minimum degree
at least three.
Let $v$ be an arbitrary vertex of $G$. We will prove that its final
charge $\ch^*(v)$
is non-negative. In order to do so, we consider a few cases regarding
its degree.
So, suppose first that $v$ is a $3$-vertex. If $v$ is a safe vertex,
then by Rule R1 its final charge is $\ch^*(v)=-1 + 3\cdot \frac{1}{3}=0$. Similarly, if 
$v$ is dangerous, then $\ch^*(v)=-1 + 2\cdot \frac{1}{2}=0$.
If $v$ is a $4$-vertex then it neither receives nor sends any charge. Thus,
$\ch^*(v)=\ch(v)=0$.

Finally, suppose that $v$ is of degree $d\ge 5$. Notice that $v$ may send
charge only by Rule R5. This may occur at most
$d/2$ times if $d$ is even, and at most $\lfloor d/2\rfloor-1$ times if
$d$ is odd (since two $3$-faces are not adjacent). Thus, 
$\ch^*(v)\ge d -4 - \left\lfloor \frac{d}{2}\right\rfloor\cdot \frac{2}{3}$, which is
non-negative if $d\ge6$. For $d=5$, $\ch^*(v)\ge5-4-\frac{2}{3}>0$.

\paragraph{Final charge of faces.}

Let $f$ be an arbitrary face of $G$.
Denote by $\x$ and $\z$ the number of
$3$-faces and the number of bad faces adjacent to $f$, respectively.
Denote by $\y$ and $\dang$ the number of safe vertices and
the number of dangerous vertices incident
to $f$, respectively.
We will prove that the final
charge $\ch^*(f)$ of $f$
is non-negative. In order to do so, we consider a few cases regarding
the size of $f$.

\paragraph{$f$ is a $3$-face.}

It is adjacent
only to $(\ge7)$-faces by Lemma~\ref{lem:kms}$(iii)$. Thus, by Rule R2, $f$ receives $1/3$ from each of its
three adjacent faces, so we obtain $\ch^*(f)=0$.

\paragraph{$f$ is a $4$-face.} It neither receives nor sends any charge. Thus,
$\ch^*(f)=\ch(f)=0$.

\paragraph{$f$ is a $5$-face.} Then, $f$ is adjacent only to
$(\ge5)$-faces due to Lemma~\ref{lem:kms}$(iii)$. So a $5$-face may send
charge only to its incident
$3$-vertices, which are all safe. Consider the following cases regarding the
number $\y$ of such vertices.

\begin{description}
\item{$\y\le3$:} Then,
$\ch^*(v) \ge 1 - 3\cdot \frac{1}{3}=0$.
\item{$\y=4$:} In this case, $f$ is a bad face.
 According to
Corollary~\ref{cor}$(vii)$,
at least two of the faces that are adjacent to $f$ have size at least $7$. Thus,
according to Rule R3, $f$ receives $1/6$ from at least two of its adjacent
faces. Hence, we conclude that
$\ch^*(v) \ge 1 - 4\cdot \frac{1}{3} +  2\cdot \frac{1}{6} =0$.
\item{$\y=5$:} Then $f$ is a very-bad face, and so,
according to Corollary~\ref{cor}$(vi)$, at least three faces adjacent to
$f$ have size at least $7$. Moreover, all faces adjacent to $f$ have
size at least $6$ by Lemma~\ref{lem:kms}$(iii)$ and
Corollary~\ref{cor}$(i)$.
By Rules R3 and R4, it follows that
the neighbouring faces of $f$ send at least $4 \cdot 1/6$ to $f$,
which implies that
$\ch^*(v) \ge 1 - 5\cdot \frac{1}{3} +  4\cdot \frac{1}{6} =0$. 
\end{description}

\paragraph{$f$ is a $6$-face.}

By Lemma~\ref{lem:kms}$(iii)$, $\x=0$.
Denote by $\zz$ number of very-bad faces adjacent to $f$.
The final charge of $f$ is
$2-{\dang}\cdot\frac{1}{2}-{\y\cdot\frac{1}{3}}-{\zz\cdot\frac{1}{12}}$
due to Rules R1 and R4.

According to Corollary~\ref{cor}$(iv)$, two dangerous vertices on $f$
cannot be adjacent so there are at most three dangerous vertices on $f$.
Observe also that $\zz\le \y/2$
by Corollary~\ref{cor}$(i)$ and because a very-bad face adjacent to $f$ is
incident to two safe vertices of $f$.
Let us consider the final charge of $f$ regarding its number of dangerous
vertices.

\begin{description}

  \item{$\dang=3$:} Since a safe vertex is not incident to a
  $(\le4)$-face, there is at most one
  safe vertex incident to $f$, i.e. $\y\le1$. Thus, $\zz=0$, and hence,
  $\ch^*(f)\ge2-3\cdot\frac{1}{2}-\frac{1}{3}>0$.

  \item{$\dang=2$:} Then, $\y\le3$.
  Let us distinguish two cases according to the value of $\y$.

  \begin{description}
    \item{$\y=3$:} Notice that $\zz=0$, otherwise it would
    contradict the reducibility of (L3).
    Hence, $\ch^*(f)\ge2-2\cdot\frac{1}{2}-3\cdot\frac{1}{3}=0$.

    \item{$\y\le2$:} In this case, there is at
    most one very-bad face adjacent to $f$, so
    $\ch^*(f)\ge2-2\cdot\frac{1}{2}-2\cdot\frac{1}{3}-\frac{1}{12}>0$.
  \end{description}
  
  \item{$\dang=1$:} Then, $\y\le4$ and $\zz\le1$
  because (L3) is reducible. So,
  $\ch^*(f)\ge2-\frac{1}{2}-\frac{4}{3}-\frac{1}{12}>0$.

  \item{$\dang=0$:} If $\y\ge5$ then, because (L3) is reducible, $\zz=0$,
  therefore $\ch^*(f)\ge2-\frac{6}{3}=0$. And, if $\y\le4$, then
  $\zz\le2$, so $\ch^*(f)\ge2-4\cdot\frac{1}{3}-2\cdot\frac{1}{12}>0$.
\end{description}

\paragraph{$f$ is a $7$-face.}

The final charge of $f$ is at least
$3-{\dang}\cdot\frac{1}{2}-{(\x+\y)\cdot\frac{1}{3}}-{\z\cdot\frac{1}{6}}$.

According to Corollary~\ref{cor}$(v)$, four dangerous vertices cannot be consecutive
on $f$, hence there cannot be more than five
dangerous vertices on $f$. Denote by
$\alpha_1,\alpha_2,\ldots,\alpha_7$ the vertices of $f$ in clockwise
order. Let ${\mathcal D}$ be the set of dangerous vertices of $f$,
so $\dang=|{\mathcal D}|$.
We shall look at the final charge of
$f$, regarding its number \dang of dangerous vertices.

\begin{description}

\item{$\dang=5$:} Up to symmetry,
${\mathcal D}=\{\alpha_1,\alpha_2,\alpha_3,\alpha_5,\alpha_6\}$.
Suppose first that $\alpha_5$ and $\alpha_6$ are not incident to a
same $(\le4)$-face. Then, there can be neither a safe vertex incident
to $f$ nor a bad face adjacent to $f$,
because a safe vertex is not incident to a $(\le 4)$-face, and also a bad face
is not adjacent to a $(\le 4)$-face. Moreover, by
Corollary~\ref{cor}$(iii)$, there is no $3$-face adjacent to $f$. Therefore,
$\ch^*(f)\ge 3-\frac{5}{2}>0$. Now, if $\alpha_5$ and $\alpha_6$ are
incident to a same $(\le4)$-face, then the
vertex $\alpha_4$ must be a $(\ge4)$-vertex by the
reducibility of (L7), and because it is not a dangerous vertex. Hence, there
is no safe vertex and no bad face adjacent to $f$, so its charge is
$\ch^*(f)\ge3-\frac{5}{2}-\frac{1}{3}>0$.

\item{$\dang=4$:} We consider several
subcases, according to the relative position of the dangerous vertices on
$f$. Recall that, by Corollary~\ref{cor}$(v)$, there are at most three
consecutive dangerous vertices. Without loss of generality, we only need
to consider the following three possibilities:

  \begin{description}
  \item ${\mathcal D}=\{\alpha_1,\alpha_2,\alpha_3,\alpha_5\}$:
  The charge of $f$ is
  $\ch^*(f)=1-(\x+\y)\cdot\frac{1}{3}-\z\cdot\frac{1}{6}$. Moreover,
  $\y\le2$, $\z\le1$ and $\x+\y\le3$ by Corollary~\ref{cor}$(iii)$
  and because a safe vertex is
  not incident to
  a $(\le4)$-face. So, $\ch^*(f)$ is negative if and only if $\y=2, \z=1$
  and $\x=1$. But in this case, the obtained configuration is (L8),
  which is reducible.

  \item ${\mathcal D}=\{\alpha_1,\alpha_2,\alpha_4,\alpha_5\}$: As a
  bad face is neither adjacent to a $(\le4)$-face nor incident to a dangerous
  vertex, we get $\z\le1$. Observe also that, as $\alpha_3$ is not
  dangerous, it has degree at least four by the reducibility of (L7) and
  (L11). Thus, $\y\le2$.
  Suppose first that $\z=1$, then $\y$ is one or two. According to the
  reducibility of (L10),
  we infer $\y+\x\le2$. Hence,
  $\ch^*(f)\ge3-4\cdot\frac{1}{2}-2\cdot\frac{1}{3}-\frac{1}{6}>0$.
  Suppose now that $\z=0$.
  We have $\x\le3$ and $\y\le2$.
  If $\x=3$ then $\y=0$, and
  if $\x=2$, then $\y\le1$ according to the reducibility
  of (L12).
  So, $\x+\y\le3$.
  Therefore, $\ch^*(f)\ge3-4\cdot\frac{1}{2}-(\x+\y)\cdot\frac{1}{3}\ge0$.
  
  \item ${\mathcal D}=\{\alpha_1,\alpha_2,\alpha_4,\alpha_6\}$:
  In this case, there is no bad face adjacent to $f$.
  Furthermore, by Corollary~\ref{cor}$(iii)$, $\x\le3$ and $\y\le2$,
  as the dangerous vertices $\alpha_4$ and $\alpha_6$ prevent at least
  one non-dangerous vertex from
  being safe.
  Observe that $\x+\y\neq5$ since otherwise it would contradict
  the reducibility of (L13).
  According to the reducibility of (L13), if $\x+\y=4$ then $\x=3$
  and no two $3$-faces have a common vertex. Hence, the obtained
  configuration is isomorphic to
  (L14) or (L15), which are both reducible.
  So, $\x+\y\le3$ and thus $\ch^*(f)\ge3-2-(\x+\y)\cdot\frac{1}{3}\ge0$.
  \end{description}

\item{$\dang=3$:} Again, we consider several subcases according to
the relative position of the dangerous vertices on $f$.

  \begin{description}

  \item ${\mathcal D}=\{\alpha_1,\alpha_2,\alpha_3\}$: Then $\x+\y\le3$ by
  Corollary~\ref{cor}$(iii)$, and $\z\le2$.
  Thus,
  $\ch^*(f)\ge3-{3\cdot\frac{1}{2}}-3\cdot\frac{1}{3}-2\cdot\frac{1}{6}>0$.

    \item ${\mathcal D}=\{\alpha_1,\alpha_2,\alpha_4\}$: Then,
    $\x\le4$.
    We shall now examine the situation according to each
    possible value of $\x$.

      \begin{description}
        
        \item{$\x=4$:} Necessarily, $\y\le1$ and $\z=0$.
        Now, if $\y=0$, then $\ch^*(f)\ge3-3\cdot\frac{1}{2}-4\cdot\frac{1}{3}>0$.
        And, if $\y=1$, then the safe
        vertex must be $\alpha_3$. Moreover, $\alpha_5$ must be
        a $(\ge5)$-vertex because (L9) is reducible.
        Hence, $f$ is incident to $\alpha_5$ between two
        $3$-faces, so by Rule R5 the vertex $\alpha_5$ gives $\frac{2}{3}$ to $f$.
        Thus,
        $\ch^*(f)\ge3-3\cdot\frac{1}{2}-5\cdot\frac{1}{3}+\frac{2}{3}>0$.

        \item{$\x=3$:} Suppose first that one of the dangerous vertices is
        incident to a $4$-face. Necessarily, $\y\le1$ and $\z\le1$.
        Thus,
        $\ch^*(f)\ge3-{3\cdot\frac{1}{2}}-{4\cdot\frac{1}{3}}-\frac{1}{12}=0$.

        Suppose now that no dangerous vertex is incident to a $4$-face.
        In particular, $\y\le2$. If $\y=2$ then the obtained configuration
        contradicts the reducibility of (L19). Hence, $\y\le1$ and $\z\le1$.
        Therefore,
        $\ch^*(f)\ge3-3\cdot\frac{1}{2}-4\cdot\frac{1}{3}-\frac{1}{6}=0$.
        
        \item{$\x=2$:} We shall prove that $\y\le2$. This is clear
        if $\alpha_1$ and $\alpha_2$ are not incident to a same $3$-face.
        So, we may assume that the edge $\alpha_1\alpha_2$ lies on a
        $3$-face.
        But then we obtain the inequality due to the reducibility of
        (L19) and (L20).
        Using Corollary~\ref{cor}$(i)$ and $\y\le2$, we infer that $\z\le1$.
         Hence, $\ch^*(f)\ge3-{3\cdot\frac{1}{2}}-4\cdot\frac{1}{3}-\frac{1}{6}=0$.

        \item{$\x=1$:} Then $\y\le3$ and $\z\le2$. If $\y=3$ and $\z=2$,
        the obtained configuration contradicts the reducibility (L20)
        or (L21).
        So, $\ch^*(f)\ge3-3\cdot\frac{1}{2}-4\cdot\frac{1}{3}-\frac{1}{6}=0$.

        \item{$\x=0$:} Again, $\y\le3$ and $\z\le2$, so
        $\ch^*(f)\ge3-3\cdot\frac{1}{2}-3\cdot\frac{1}{3}-2\cdot\frac{1}{6}>0$.
        
      \end{description}

\item ${\mathcal D}=\{\alpha_1,\alpha_2,\alpha_5\}$: As in the previous
case, $\x\le4$ and we look at all the possible cases according to the
    value of $\x$. Since a bad face is not incident to a dangerous
    vertex, notice that only edges $\alpha_3\alpha_4$ and
    $\alpha_6\alpha_7$ can be incident to a bad face. In particular,
    $\z\le2$.
      \begin{description}

        \item{$\x=4$:} In this case, $\y=0$ and $\z=0$. Therefore,
        $\ch^*(f)=3-3\cdot\frac{1}{2}-4\cdot\frac{1}{3}>0$.

        \item{$\x=3$:} If one of the dangerous vertices is
        incident to a $4$-face then $\y=0$, hence $\z=0$.
        Thus, $\ch^*(f)\ge3-3\cdot\frac{1}{2}-3\cdot\frac{1}{3}\ge0$.
        So now, we infer that $\y$ cannot be $2$, otherwise it would contradict
        the reducibility of (L16). Therefore, $\y$ is at most one, and
        so $\z\le1$ by Corollary~\ref{cor}$(i)$. Thus,
        $\ch^*(f)\ge3-3\cdot\frac{1}{2}-4\cdot\frac{1}{3}-\frac{1}{6}=0$.

        \item{$\x=2$:} According to the reducibility of (L16) and (L17),
        $\y\le2$. As
        $\ch^*(f)=3-{3\cdot\frac{1}{2}}-{(\x+\y)\cdot\frac{1}{3}}-{\z\cdot\frac{1}{6}}$,
        we deduce
        $\ch^*(f)<0$ if and only if $\y=2$ and $\z=2$. In this case, the
        obtained configuration is (L18), which is reducible.

        \item{$\x=1$:} Because (L16) and (L17) are reducible,
        $\y\le2$. So,
        $\ch^*(f)\ge3-3\cdot\frac{1}{2}-{3\cdot\frac{1}{3}}-{2\cdot\frac{1}{6}}>0$.

        \item{$\x=0$:} Then $\y\le3$, and so
        $\ch^*(f)\ge3-3\cdot\frac{3}{2}-3\cdot\frac{1}{3}-2\cdot\frac{1}{6}>0$.
      \end{description}

      \item{${\mathcal D}=\{\alpha_1,\alpha_3,\alpha_5\}$:} In this
      case, $\y\le2$ since a safe vertex is not incident to a
      $(\le4)$-face,
      and $\z\le1$, since a bad face cannot be incident to a dangerous
      vertex. Moreover, $\x\le4$. Let us examine the possible cases
      regarding the value of $\x$.

      \begin{description}
        \item{$\x=4$:} Observe that $\y\le1$ and $\z=0$.
        Note also one of $\alpha_2,\alpha_3,\alpha_6,\alpha_7$ is
        adjacent to a dangerous vertex, and incident to $f$ between
        two triangles. Hence, by the reducibility of (L9), it has
        degree at least five, and by Rule R5, it sends $\frac{2}{3}$
        to $f$. Thus,
        $\ch^*(f)\ge3-{3\cdot\frac{1}{2}}-{5\cdot\frac{1}{3}}
        +\frac{2}{3}>0$.

        \item{$\x=3$:}
        If $\y\le1$ then
        $\ch^*(f)\ge3-3\cdot\frac{1}{2}-4\cdot\frac{1}{3}-\frac{1}{6}=0$.
        And, if $\y=2$ then, up to symmetry,
        the two safe vertices are either $\alpha_6$ and $\alpha_7$, or
        $\alpha_2$ and $\alpha_6$. In the former case, one of
        $\alpha_2,\alpha_4$ is
        incident to $f$ at the intersection of two $3$-faces.
        Furthermore, it
        must be a $(\ge5)$-vertex due to the reducibility of (L9).
        In the latter case, the same holds for $\alpha_4$ due to
        the reducibility of (L9). Hence, in both cases
        the face $f$ receives $2/3$ from one of its incident vertices
        by Rule R5. Recall that $\z\le1$, and therefore,
        $\ch^*(f)\ge3-3\cdot\frac{1}{2}-{5\cdot\frac{1}{3}}-{2\cdot\frac{1}{6}}+\frac{2}{3}>0$.

        \item{$\x\le2$:} As $\y\le2$ and $\z\le1$, we infer that
        $\ch^*(f)\ge3-3\cdot\frac{1}{2}-4\cdot\frac{1}{3}-\frac{1}{6}=0$.
        
      \end{description}
    \end{description}

\item{$\dang=2$:} Again, we consider several subcases, regarding the
position of the dangerous vertices on $f$.

  \begin{description}
    
    \item ${\mathcal D}=\{\alpha_1,\alpha_2\}$: Observe that $\z\le3$,
    and according to Corollary~\ref{cor}$(iii)$,
    $\x+\y\le6$. We consider three cases, according to the value of $\x+\y$.

      \begin{description}
        \item{$\x+\y=6$:} All the vertices incident to $f$ have degree
        three, and $f$ is adjacent to a $3$-face. Thus, by
         Corollary~\ref{cor}$(ii)$, $f$  is not adjacent to
         any $(\le6)$-face. In particular, no bad face is adjacent to
         $f$, i.e. $\z=0$. Hence, $\ch^*(f)\ge3-1-6\cdot\frac{1}{3}=0$.

        \item{$\x+\y=5$:} If $\z\le2$, then
        $\ch^*(f)\ge3-1-5\cdot\frac{1}{3}-2\cdot\frac{1}{6}=0$.
        Otherwise, $\z=3$. Note that the edge $\alpha_1\alpha_2$
        must be incident to a $(\le4)$-face.
        If this face is of size four, then we obtain configuration
        (L22). Suppose now that this face is of size three.
        Since there is no three consecutive bad faces around $f$,
        we can assume that each of the edges $\alpha_3\alpha_4$ and
        $\alpha_6\alpha_7$ lies on a bad face.
        By the reducibility of (L18), we conclude that $\alpha_3$
        and $\alpha_7$ have degree at least four. But then, $\x+\y<5$.

        \item{$\x+\y\le4$:} In this case,
        $\ch^*(f)\ge3-1-4\cdot\frac{1}{3}-3\cdot\frac{1}{6}>0$.
      \end{description}
    
    \item{\it ${\mathcal D}=\{\alpha_1,\alpha_3\}$ or ${\mathcal
    D}=\{\alpha_1,\alpha_4\}$}: Again $\x+\y\le6$, and we consider two cases
    regarding the value of $\x+\y$. Since a bad face
    is not incident to a dangerous vertex, we infer that $\z\le3$.

    \begin{description}

      \item{$\x+\y=6$:} Suppose first that ${\mathcal
      D}=\{\alpha_1,\alpha_3\}$. Let $P_1=\alpha_1\alpha_2\alpha_3$ and
      $P_2=\alpha_3\alpha_4\alpha_5\alpha_6\alpha_7\alpha_1$. In order to assure
      $\x+\y=6$, observe that all edges of $P_1$ are incident to
      $3$-faces and all inner vertices of $P_2$ are safe, or vice-versa.
      Thus, $\alpha_2$ or $\alpha_4$ is a $(\ge5)$-vertex by the
      reducibility of (L9). Hence, it gives $\frac{2}{3}$ to $f$
      by Rule R5.
      Therefore, 
      $\ch^*(f)\ge3-2\cdot\frac{1}{2}-6\cdot\frac{1}{3}-3\cdot\frac{1}{6}+\frac{2}{3}>0$.
     
      Suppose now that ${\mathcal D}=\{\alpha_1,\alpha_4\}$. Similarly as above,
      one can show that $\alpha_2$ or $\alpha_5$ is a $(\ge5)$-vertex
      that donates $\frac{2}{3}$ to $f$.
      Hence,
      $\ch^*(f)\ge3-2\cdot\frac{1}{2}-6\cdot\frac{1}{3}-\frac{3}{6}+\frac{2}{3}>0$.

      \item{$\x+\y\le5$:} Notice that $\z\le2$.
      Therefore, $\ch^*(f)\ge3-2\cdot\frac{1}{2}-{5\cdot\frac{1}{3}}-2\cdot\frac{1}{6}=0$.

    \end{description}

  \end{description}

\item{$\dang=1$:} Then $\x+\y\le6$ and, by Corollary~\ref{cor}$(i)$,
we infer that $\z\le3$.
So, $\ch^*(f)\ge3-\frac{1}{2}-6\cdot\frac{1}{3}-3\cdot\frac{1}{6}=0$.

\item{$\dang=0$:} By Corollary~\ref{cor}$(i)$, $\x+\y\le7$ and $\z\le4$.
So, $\ch^*(f)\ge3-7\cdot\frac{1}{3}-4\cdot\frac{1}{6}=0$.

\end{description}

\paragraph{$f$ is an $8$-face.}

  Because (L4) and (L23) are reducible, there cannot be three consecutive
  dangerous vertices on $f$. Hence, $\dang\le5$.
  Denote by $\alpha_i$, $i\in\{1,2,\ldots,8\}$, the
  vertices incident to $f$ in clockwise order, and let ${\mathcal D}$ be
  the set of dangerous vertices incident to $f$.

  \begin{description}

    \item{$\dang=5$:} Up to symmetry, ${\mathcal
    D}=\{\alpha_1,\alpha_2,\alpha_4,\alpha_5,\alpha_7\}$.
    Since a bad face is not incident to a dangerous
    vertex, necessarily $\z=0$. For $i\in\{1,4\}$, denote by $f_i$ the face
    adjacent to $f$ and incident to
    both $\alpha_i$ and $\alpha_{i+1}$.
    Since (L24) is reducible, at most one of $f_1$
    and $f_4$ is a
    $3$-face. Furthermore, at most two of $\alpha_3,\alpha_6,\alpha_8$ can be safe
    vertices, since at least one of $\alpha_6,\alpha_8$ is a
    $(\ge4)$-vertex. Therefore, $\x\le2$,
    $\y\le2$ and so, $\ch^*(f)\ge4-5\cdot\frac{1}{2}-4\cdot\frac{1}{3}>0$.

    \item{$\dang=4$:} Up to symmetry, the set of dangerous vertices
    is $\{\alpha_1,\alpha_2,\alpha_4,\alpha_5\}$,
    $\{\alpha_1,\alpha_2,\alpha_5,\alpha_6\}$,
    $\{\alpha_1,\alpha_2,\alpha_4,\alpha_6\}$,
    $\{\alpha_1,\alpha_2,\alpha_4,\alpha_7\}$ or
    $\{\alpha_1,\alpha_3,\alpha_5,\alpha_7\}$.
    In any case, $\z\le2$ and $\x+\y\le5$. Hence, $\ch^*(f)\ge
    4-\frac{4}{2}-\frac{5}{3}-\frac{2}{6}=0$.

    \item{$\dang=3$:} Then,
    $\x+\y\le6$ and $\z\le3$. So,
    $\ch^*(f)\ge4-\frac{3}{2}-\frac{6}{3}-\frac{3}{6}=0$.

    \item{$\dang=2$:} Then,
    $\x+\y\le7$, and by Corollary~\ref{cor}$(i)$, $\z\le4$. Thus,
    $\ch^*(f)\ge4-\frac{2}{2}-\frac{7}{3}-\frac{4}{6}=0$.

    \item{$\dang=1$:} Again, $\x+\y\le7$ and
    $\z\le4$, so $\ch^*(f)\ge4-\frac{1}{2}-\frac{7}{3}-\frac{4}{6}>0$.

    \item{$\dang=0$:} By Corollary~\ref{cor}$(i)$, $\z\le5$. So,
    $\ch^*(f)\le4-\frac{8}{3}-\frac{5}{6}>0$.
    
  \end{description}

\paragraph{$f$ is a $(\ge9)$-face.}


Let $f$ be a $k$-face with $k\ge9$, and denote by $u_1,u_2,\ldots,u_\dang$
the dangerous vertices on $f$ in clockwise order.
Denote by $f_i$ the $(\le4)$-face incident to $u_i$.
The facial segment $P=u_iw_1w_2\ldots w_ju_{i+1}$ of $f$ between
$u_i$ and $u_{i+1}$ (in clockwise order) is of one of the five following
types: 
\begin{description}
\item[$(a)$]
if $j\ge1$, $w_1$ is not incident to $f_i$ and $w_j$ is not incident to
$f_{i+1}$;
\item[$(b)$] if $j\ge1$, $w_1$ is incident to $f_i$ and $w_j$
is incident to $f_{i+1}$;
\item[$(c)$] if $j\ge1$ and not of type $(a)$ or $(b)$;
\item[$(d)$] if $j=0$ and both $f_i$ and $f_{i+1}$ are the
same $3$-face; and
\item[$(e)$] if $j=0$ and not of type $(d)$.
\end{description}

We denote by $\alpha$ the number
of paths of type $(a)$, $\beta$ the number of paths of type $(b)$,
$\gamma$ the number of paths of type $(c)$, $\delta$ the number of paths
of type $(d)$ and $\varepsilon$ the number of paths of type $(e)$. Note
that a path of type $(d)$ or $(e)$ is of length one.
Observe that the following holds:
\begin{claim}\label{c1}
$\alpha+\beta+\gamma+\delta+\varepsilon=\dang$.
\end{claim}
\noindent
We now bound the number of safe vertices and $3$-faces.
\begin{claim}\label{c2}
$\x+\y\le k -\alpha-\gamma-\varepsilon$.
\end{claim}
For each \l-path $P$ of
type $(a), (c)$ or $(e)$
the number of safe vertices on $P$ plus the number of
$3$-faces which share an edge with $P$ is at most $\l-1$. Indeed,
for any path of one of these types, there are at most \l faces
different from $f$ and incident to an edge of the path, but at least one of them
is not a $(\le4)$-face. There are $\l-1$ vertices on the path, so at most
$\l-1$ safe vertices. Furthermore, every $(\le4)$-face prevents at least one
vertex from being safe. Observe also that an \l-path of type $(b)$ or $(d)$
contributes for at most \l, which thus yields Claim~\ref{c2}.

\medskip
We distinguish two kinds of paths of type $(e)$: a path of type $(e)$
is of {\it type $(e_0)$} if its edge
is not incident to a $4$-face. Otherwise, it is of {\it type $(e_1)$}.
Let $\varepsilon_i$ be the number of paths of type $(e_i)$,
$i\in\{0,1\}$.

\begin{claim}\label{c3}
$\z\le k-2\dang+\delta+\varepsilon_1$.
\end{claim}
\noindent
First, remark that
each dangerous vertex prevents its two incident edges on $f$ from
belonging to a bad face, since no bad face is incident to a dangerous
vertex. By the reducibility of
(L23), there cannot be three consecutive dangerous vertices on
$f$, so it only remains to consider two consecutive dangerous vertices,
i.e. paths of type $(d)$ or $(e)$.
A path of type $(d)$ or $(e_1)$ prevents exactly three edges of $f$ from being
incident to a bad face. Every $1$-path of type $(e_0)$ prevents
at least four edges of $f$ from being incident to a bad face. To see
this, consider a path $u_1u_2u_3u_4u_5u_6$, where $u_2u_3$ is a $1$-path
of type $(e_0)$. Clearly, none of $u_1u_2,u_2u_3,u_3u_4$ is incident
to a bad face. We claim that at least one of
$u_4u_5,u_5u_6$ is not incident to a bad face. Otherwise, if $u_4u_5$
is incident to a bad face, then by Lemma~\ref{lem:kms}$(iii)$,
$u_4$ must be a $(\ge4)$-vertex.
Hence, by Corollary~\ref{cor}$(i)$, $u_5u_6$ is not incident to
a bad face.
As no
three dangerous vertices are consecutive of $f$, this
proves Claim~\ref{c3}.

\medskip
\begin{claim}\label{c4}
$\alpha-\beta+\varepsilon_0=\delta+\varepsilon_1$.
\end{claim}
\noindent
Associate each dangerous
vertex $u_i$ with its incident $(\le4)$-face $f_i$. Each path
of type $(a)$ contains no face $f_i$, so does each path
of type $(e_0)$; each path of type $(c)$ contains
exactly one face $f_i$, and each path of type $(b),(d)$ or $(e_1)$ contains
exactly two faces $f_i$ (where a face is counted with its multiplicity,
i.e. once for each dangerous vertex of $f$ incident to it). So,
$\dang=\gamma+2(\beta+\delta+\varepsilon_1)$, and hence
$\alpha+\beta+\gamma+\delta+\varepsilon=\gamma+2(\beta+\delta+\varepsilon_1)$,
which gives Claim~\ref{c4}.

\medskip
So, by Claims~\ref{c1}--\ref{c4}, we get
\begin{eqnarray*}
\ch^*(f)&=&k-4-\dang\cdot\frac{1}{2}-(\x+\y)\cdot\frac{1}{3}-\z\cdot\frac{1}{6}\\
~&\ge&k-4-\frac{\dang}{2}-\frac{k-\alpha-\gamma-\varepsilon}{3}-\frac{k-2\dang+\delta+\varepsilon_1}{6}\\
~&=&
\frac{k}{2}-4-\frac{\dang}{6}+\frac{\alpha+\gamma+\varepsilon_0}{3}+\frac{\varepsilon_1-\delta}{6}\\
~&=&\frac{k}{2}-4+\frac{(\alpha-\beta+\varepsilon_0)+\gamma}{6}-\frac{\delta}{3}\\
~&=&\frac{k}{2}-4+\frac{\delta+\varepsilon_1+\gamma}{6}-\frac{\delta}{3}\\
~&\ge&\frac{k}{2}-4-\frac{\delta}{6}.
\end{eqnarray*}

According to Corollary~\ref{cor}$(iii)$ and the reducibility of (L24),
there are at least two vertices
between any two paths of type $(d)$. So,
$\delta\le\frac{k}{4}$.
Therefore, one can conclude that
\[
\ch^*(f)\ge\frac{k}{2}-\frac{k}{24}-4=\frac{11}{24}k-4\ge\frac{99}{24}-4>0.
\]

\bibliographystyle{plain}
\bibliography{facial}

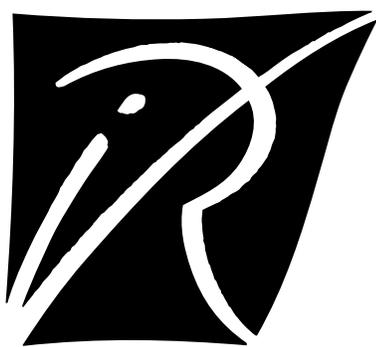
\end{document}